\documentclass[preprint2]{aastex62}
\usepackage[version=4]{mhchem}

\newcommand{\kms}{$\rm km~s^{-1}$}

\graphicspath{{./}{figures/}}

\shorttitle{Orion source I}
\shortauthors{Wright et al.}

\begin{document}

\title{Observations of Orion Source I Disk and Outflow Interface}



\author{Melvyn Wright}
\affiliation{Radio Astronomy Lab, University of California, 501 Campbell Hall, Berkeley CA 94720-3441, USA}
\author[0000-0001-6765-9609]{Richard Plambeck}
\affiliation{Radio Astronomy Lab, University of California, 501 Campbell Hall, Berkeley CA 94720-3441, USA}
\author{Tomoya Hirota}
\affiliation{Mizusawa VLBI Observatory, National Astronomical Observatory of Japan, Osawa 2-21-1, Mitaka, Tokyo 181-8588, Japan}
\author{Adam Ginsburg}
\affil{National Radio Astronomy Observatory, Charlottesville, VA 22903, USA}
\author{Brett McGuire}
\affil{National Radio Astronomy Observatory, Charlottesville, VA 22903, USA}
\author{John Bally}
\affil{CASA, University of Colorado, 389-UCB, Boulder, CO 80309, USA}
\author{Ciriaco Goddi}
\affil{Leiden Observatory, Leiden University, P.O. Box 9513, 2300 RA Leiden, The Netherlands}

\begin{abstract}
We imaged the continuum and molecular line emission from Orion Source~I (SrcI) with up to 30 mas (12 AU) resolution at 43, 99, 223, and 340 GHz in an attempt to probe the structure and chemistry of the circumstellar disk and bipolar outflow associated with this high mass protostar.  The continuum spectral index ranges from $\sim$2 along the midplane of the disk to $\sim$3 along the edges, consistent with dust that is optically thick in the midplane but becomes optically thin at the periphery.  Salt (NaCl) emission is visible where the dust is optically thin; it provides a unique tracer of the velocity field within the disk.  All other molecules that we have mapped -- H$_2$O, AlO, SiO, SiS, SO, and SO$_2$ -- appear to originate primarily in the bipolar outflow.  The base of the outflow is corotating with the disk.  SiS shows a filamentary structure that is most prominent along the edges of the outflow.  The molecular distributions suggest that Si and Al released from dust grains in the disk react with oxygen derived from H$_2$O to form SiO and AlO, and with SO and SO$_2$ to form SiS.

\end{abstract}


%

\section{Introduction} \label{sec:intro}

The Kleinmann-Low Nebula in Orion, at a distance 415 pc   \citep{Menten2007,Kim2008,Kounkel2018}, is the nearest interstellar cloud in which massive ($M >8$~$M_{\odot}$) stars are forming. The two most massive stars in this region, Source~I (SrcI) and the Becklin-Neugebauer Object (BN), appear to be recoiling from one another at 35-40~\kms\ \citep{Rodriguez2005, Gomez2008, Goddi2011}, suggesting that they were ejected from a multiple system via dynamical decay approximately 500 years ago \citep{Bally2017}.  
SrcI has a mass $\sim$15~$M_{\odot}$ \citep{Ginsburg2018}, with a rotating accretion disk and a molecular outflow that is prominent in SiO.
The disk around SrcI has been
well studied as it is the closest known disk around a high mass protostar
\citep{Hirota2014, Plambeck2016, Ginsburg2018}; it is associated with SiO and \ce{H2O} masers \citep{Reid2007, Goddi2009, Plambeck2009, Matthews2010, Goddi2011,  Niederhofer2012, Greenhill2013}.  Recently dozens of spectral lines of NaCl and KCl were identified in this disk 
\citep{Ginsburg2019}. The rich chemistry of
SrcI's disk and outflow raises the question of whether this source is a paradigm for high-mass star formation, or an anomaly born in the unique environment created by
the SrcI/BN interaction and associated explosive outflow \citep{Bally2017}.

In this paper, we present high resolution images of SrcI that were made from JVLA and ALMA data at 43, 99, 223, and 340~GHz.  We use continuum images with 30 mas (12 AU) resolution to probe the dust opacity across the disk.  
We discuss chemical pathways that can lead to the observed molecular distributions.

\section{Observations and Data Reduction}
Table~1 provides a summary of the observations, including project codes.

\subsection{43 GHz}
The spectral setup for the 43~GHz JVLA observations included 16 wideband windows, each covering 116~MHz bandwidth and 58 spectral channels.  The JVLA wideband setup is 64 channels covering 128 MHz; the reduction script trimmed off end channels because of analog filter rolloff. The data were self-calibrated using a strong SiO $v=1$, $J=1-0$ maser feature at -3.5~\kms\  that was observed simultaneously in a narrowband window with spectral resolution 0.3 km s$^{-1}$.


The wide band average with a mean frequency of 42.65~GHz and a bandwidth of 1.4~GHz excludes 4 spectral windows containing spectral line emission. We made images with the synthesized beam width of
56~$\times$~44~mas, and convolved to 30~mas { and 50~mas resolution} for comparison with images at 99~GHz (see below).

\subsection{99 GHz}
ALMA observations at 99~GHz (Band 3; B3) on 2017 Oct 12 and 17 included 4 spectral windows, each with a bandwidth of 1.875~GHz and 960 spectral channels. The wide band average with a mean frequency of 99.275~GHz and a bandwidth of 7.5~GHz excludes channels containing spectral line emission.
{ The data were calibrated using observatory supplied scripts.} We made images with the synthesized beam width 45~$\times$~36~mas, 
and convolved to 30~mas { and 50~mas resolution} for comparison with images at 43~GHz.



\subsection{223 and 340 GHz}

The 223~GHz (ALMA Band 6; B6) and 340~GHz (ALMA Band 7; B7) observations and calibration are described in \citet{Ginsburg2018}. These data were not self-calibrated. The { observational parameters} are given in Table 1.

In order to compare the SrcI outflow with the large scale structure associated
with the SrcI/BN explosion mapped in CO by \citet{Bally2017},
we imaged other spectral lines that fell within the passband of these lower resolution observations (ADS/JAO.ALMA\#2013.1.00546.S).  The data were calibrated using the observatory-supplied CASA scripts. We made mosaic images from these data in a 30$''$ field around SrcI in  SO, SO$_2$, and SiO emission with 1.5 $\times$ 0.9 $''$ angular and 2~\kms\ velocity resolution. We used only the ALMA 12m array data in these images to filter out the large scale structure and enhance the filamentary structures.

The \textsc{Miriad} software package \citep{Sault1995} was used to image and analyze all the data in this paper.





\section{THE SRC I DISK}

We used both CLEAN and Maximum Entropy algorithms to image the JVLA and ALMA data.
We convolved the continuum images to a common resolution of 30~mas { and 50~mas resolution} in order to
make spectral index images between 43, 99, 223 and 340~GHz. Because the 43~GHz data were self-calibrated, the wideband image was spatially offset from the higher frequency images.  We aligned the 43~GHz image with the 99~GHz image using \textsc{Miriad} task {\it imdiff}
which finds optimum parameters in a maximum likelihood sense for making one image approximate another image.  The parameters adjusted were shifts { 9 mas} in RA and { -2 mas} in DEC to an image center at RA = 05:35:14.513, DEC = -05:22:30.576 used in the following figures. { The 99, 223, and 340~GHz data were not self calibrated}, only the 43~GHz image position was adjusted. Src I is moving at Dx=6.3, Dy=-4.2 mas/yr \citep{Goddi2011} . The 99, 223, and 340~GHz image data were acquired within a $\sim2$ month period (see Table 1) and the offset in the images from the proper motion is less than $\sim$0.4 AU. 

\begin{figure}
\includegraphics[width=1.0\columnwidth, clip, trim=3cm 3.7cm 2.5cm 1cm]{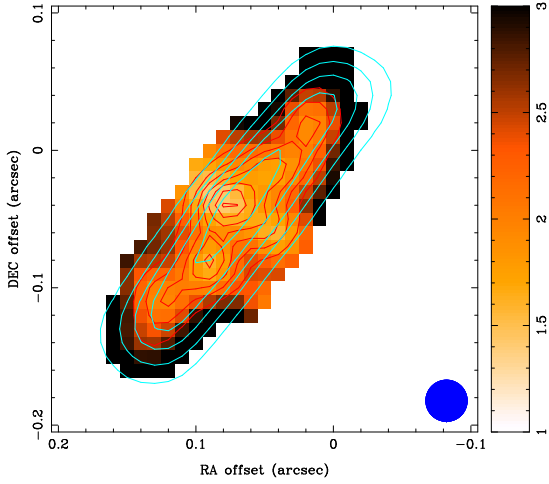}
\caption{Spectral Index from 99 to 340 GHz.
Red  contour levels at 99 GHz: 200 400 600 800 1000 1200 1400 K.
 Blue contour levels at 340 GHz: 100 200 300 400 500 600 K. The color images shows the spectral index distribution. The 30 mas convolving beam FWHM is indicated in blue in the lower right.
\label{fig:B7-B3_SI}}
\end{figure}

\begin{figure}
\includegraphics[width=1.0\columnwidth, clip, trim=3cm 3.7cm 2.5cm 1cm]{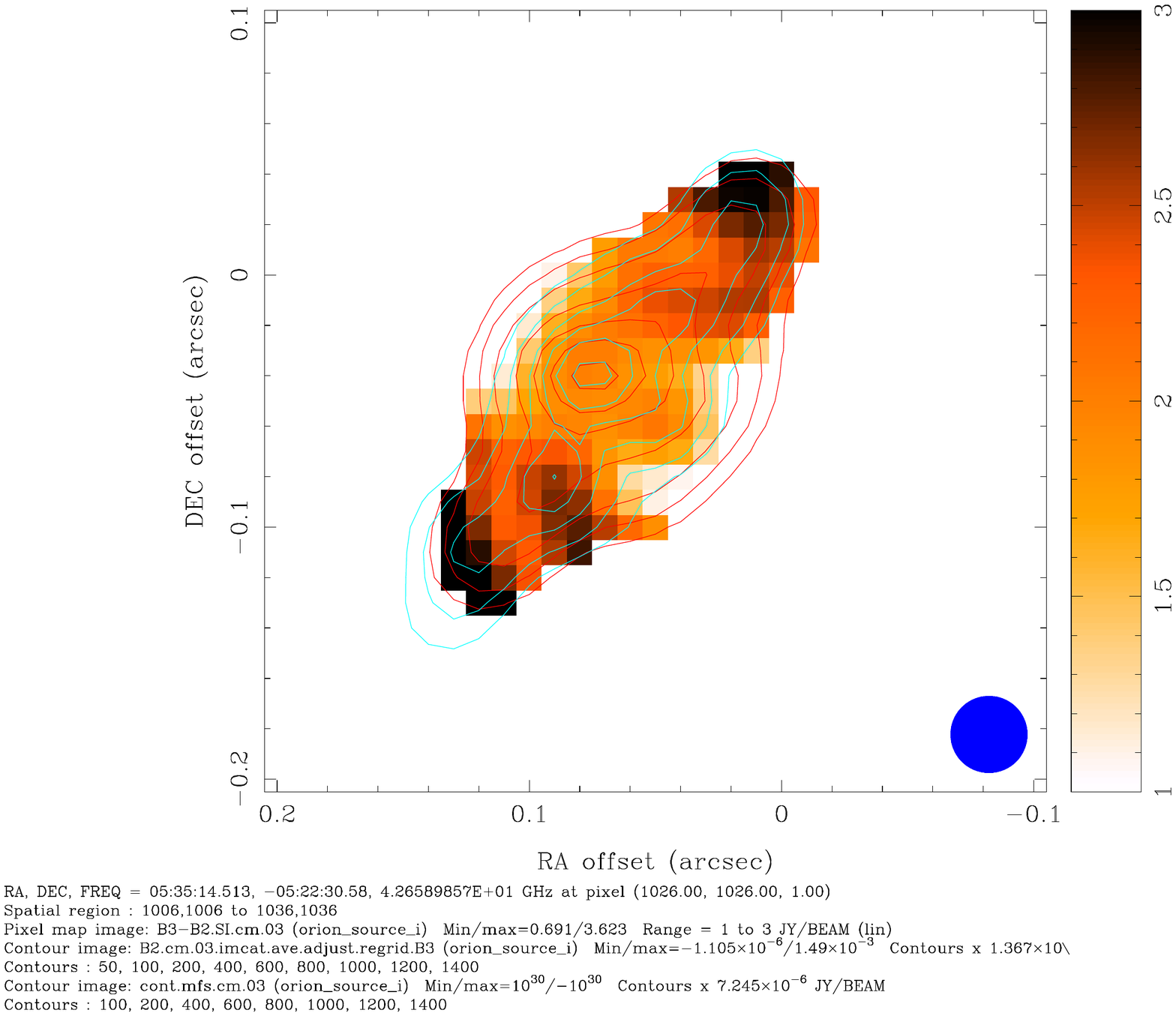}
\caption{Spectral Index 43 GHz to 99 GHz.
Red  contour levels at 43 GHz: 200 400 600 800 1000 1200 1400 K.
Blue contour levels at 99 GHz: 100 200 300 400 500 600 K. The color image shows the spectral index distribution. The 30 mas convolving beam FWHM is indicated in blue in the lower right.
\label{fig:B3-B2_SI}}
\end{figure}

\begin{figure}
\includegraphics[width=1.0\columnwidth, clip, trim=3cm 3.7cm 2.5cm 1cm]{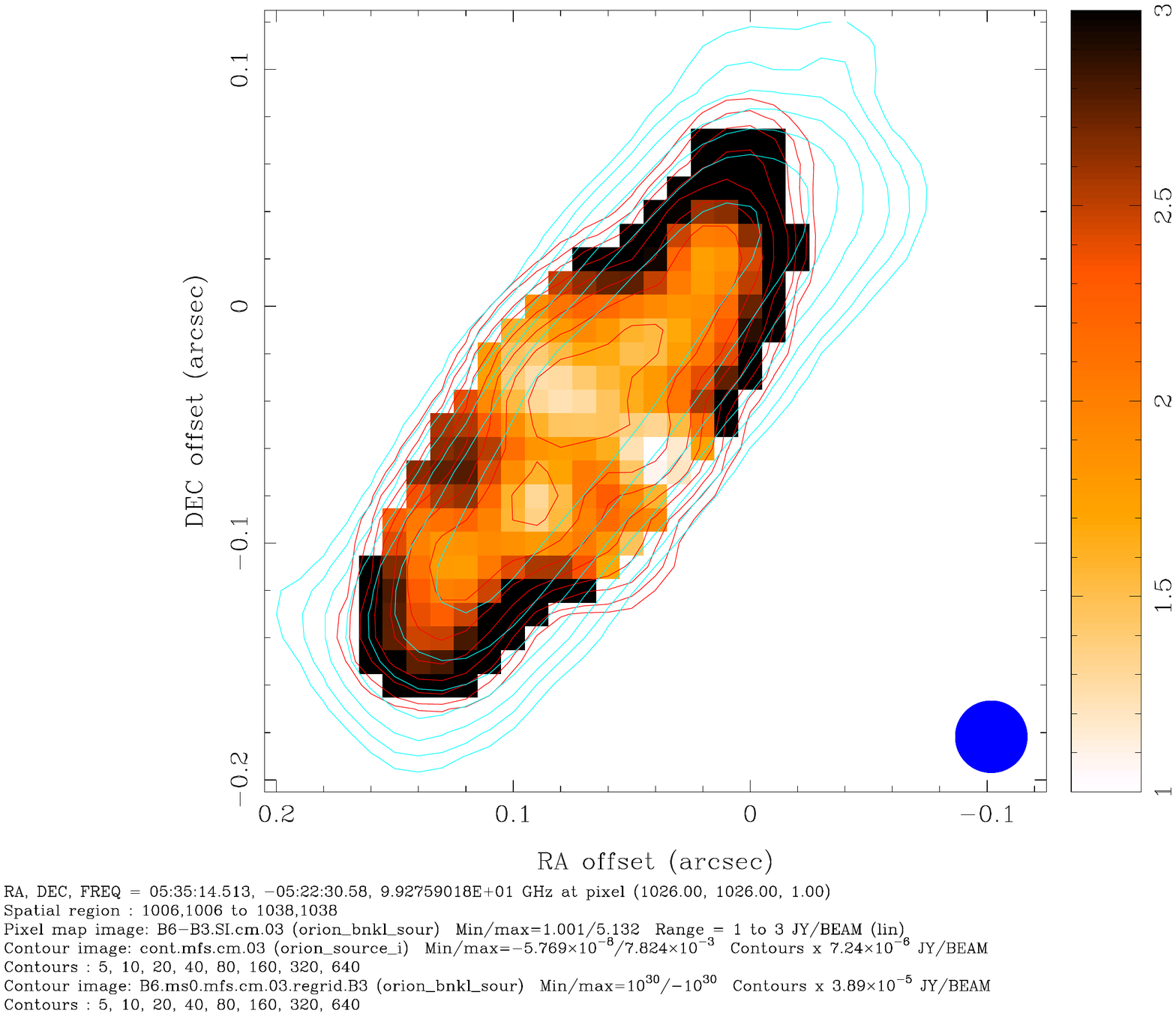}
\caption{Spectral Index from 99 to 223 GHz.
Red  contour levels at 99 GHz: 5 10 20 40 80 160 320 640 K.
Blue contour levels at 223 GHz: 5 10 20 40 80 160 320 640 K. The color image shows the spectral index distribution. The 30 mas convolving beam FWHM is indicated in blue in the lower right.
\label{fig:B6-B3_SI}}
\end{figure}

 
\subsection{Spectral Index Distribution}


Figures 1, 2 and 3  show  the spectral index computed from the ratio of continuum brightness at 340, 223, 99, and 43 GHz  { at 30~mas resolution. At 50~mas resolution, the same spectral index distributions are seen, convolved to the lower resolution}.
The RMS noise levels on the continuum images are 4~K, 3~K, 7~K, and 6~K, respectively; the images were clipped at 5\% of the peak for the spectral index calculation.  The figures show that spectral index $\sim$2 along the midplane of the disk, consistent with optically thick emission, almost certainly from dust \citep{Plambeck2016}.  The spectral index steepens to $\sim$3 at the edges and ends of the disk, indicating emission from optically thin dust.
In figure~\ref{fig:B3-B2_SI}, we see that the central source is more prominent at 43~GHz, and also note the steepening of the spectral index at the ends of the disk between 43 and 99~GHz.
   









Uncertainties in the spectral indices are dominated by the absolute flux calibration accuracy of $\sim$10\% for each frequency band.
A multiplicative error in the flux ratio is an additive error in the spectral index of +0.15 and -0.18 in the 340/99 GHz spectral index image, +0.22 and -0.28 in the 223/99 GHz image, and +0.11 and -0.13 in the 99/43 GHz image. A least squares fit from 43 to 340~Ghz to the spectral index at the central position gives 1.6 +/- 0.1, whereas at the ends of the disk  the fitted spectral index is  3.4 +/- 0.3.
The spectral index variations across the images are significant.

\subsection{Disk Structure}
   
\citet{Ginsburg2018} fitted the observed structure of the disk from B3  and B6 ALMA continuum observations at 50 and 20 mas resolution, respectively. They determined that the disk has a length of $\sim$100 AU, and  vertical FWHM height of $\sim$20 AU.  They also detected a compact source near the center of the disk, smeared parallel to the disk major axis.  
The model residuals shown by \citet{Ginsburg2018} have a halo of emission at the $\sim$30 K level in the B6 image that may be from optically thin dust, as it was not seen in the B3 image. This halo of emission is evident in Figure~\ref{fig:B6-B3_SI}, where both B3 and B6 are plotted at the same logarithmic contour levels. 
  

Table~2 summarizes the results of Gaussian fits to the disk size in our 4 continuum images.  The disk major axis increases with observing frequency, which is expected as the dust optical depth increases.  The minor axis is largest at 43 GHz, however, indicating that the central point source is more prominent at lower frequencies.

If we assume that the disk is circular and infinitesimally thin, we can set a lower limit on its inclination angle.  The 224 and 340~GHz source sizes given in Table 2 imply that this lower limit is 78-80\degr.  

\begin{figure}
\includegraphics[width=1.0\columnwidth, clip, trim=3cm 3.0cm 2.5cm 1cm]{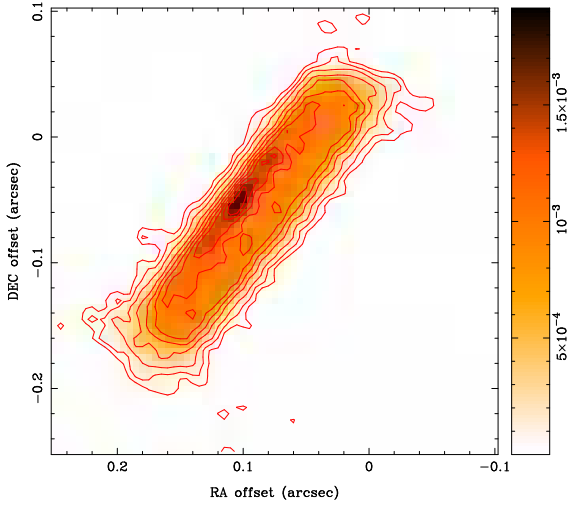}
\caption{340 GHz continuum emission. Maximum Entropy image with 5 mas pixels.
 Contours are at 25, 50, 100, 200, 300, 400, 500, 600, 700, and 800 K. The linear color scale in units Jy pixel$^{-1}$.
 (480 K mJy$^{-1}$). The peak brightness at the compact source is 874 K.
\label{fig:B7.mem}}
\end{figure}

In Figure~\ref{fig:B7.mem} we show a Maximum Entropy image of the B7 continuum emission. The lower limit to the inclination estimated from the major and minor axes of the Maximum Entropy image is 79$\pm$1$^{\circ}$ measured at the 400 K contour (99 $\times$ 19 AU), and 74$\pm$1$^{\circ}$ measured at the 25 K contour (239 $\times$ 45 AU). The lowest contours at the ends of the disk suggest a flared structure, so the inclination
measured at the 400 K contour is better determined. If the disk were inclined by as much as 74 or 79$^{\circ}$, the $\sim$ 20 AU disk thickness, would imply that the SW edge should be slightly curved.  The incredibly straight SW edge in Figure~\ref{fig:B7.mem}, suggests that the inclination is closer to 90$^{\circ}$.
   
A third estimate of the disk orientation can be obtained from the SiO outflow.
At the highest velocities, the SiO outflow is blue-shifted to the NE, and red-shifted
to the SW. This is more easily seen in Figure 1 of \citet{Plambeck2009}, which maps the large scale
structure of the SiO outflow. The blue-shifted emission at -13.5~\kms~subtends an angle 27$^{\circ}$ at a radius $\sim$1300 AU to the NE, and the red-shifted emission at 21.9~\kms~subtends an angle 26$^{\circ}$ at a radius $\sim$1400 AU. {The red-blue asymmetry in the outflow implies that the NE edge of the disk is tipped toward us. }
 Proper motion measurements of
compact structures in the outflow could enable us to derive an inclination from the SiO outflow. \citet{Matthews2010} derive an inclination $\sim$85\degr~ from SiO masers close to the disk.

\begin{figure}
\includegraphics[width=1.0\columnwidth, clip, trim=3cm 3.3cm 2.5cm 1cm]{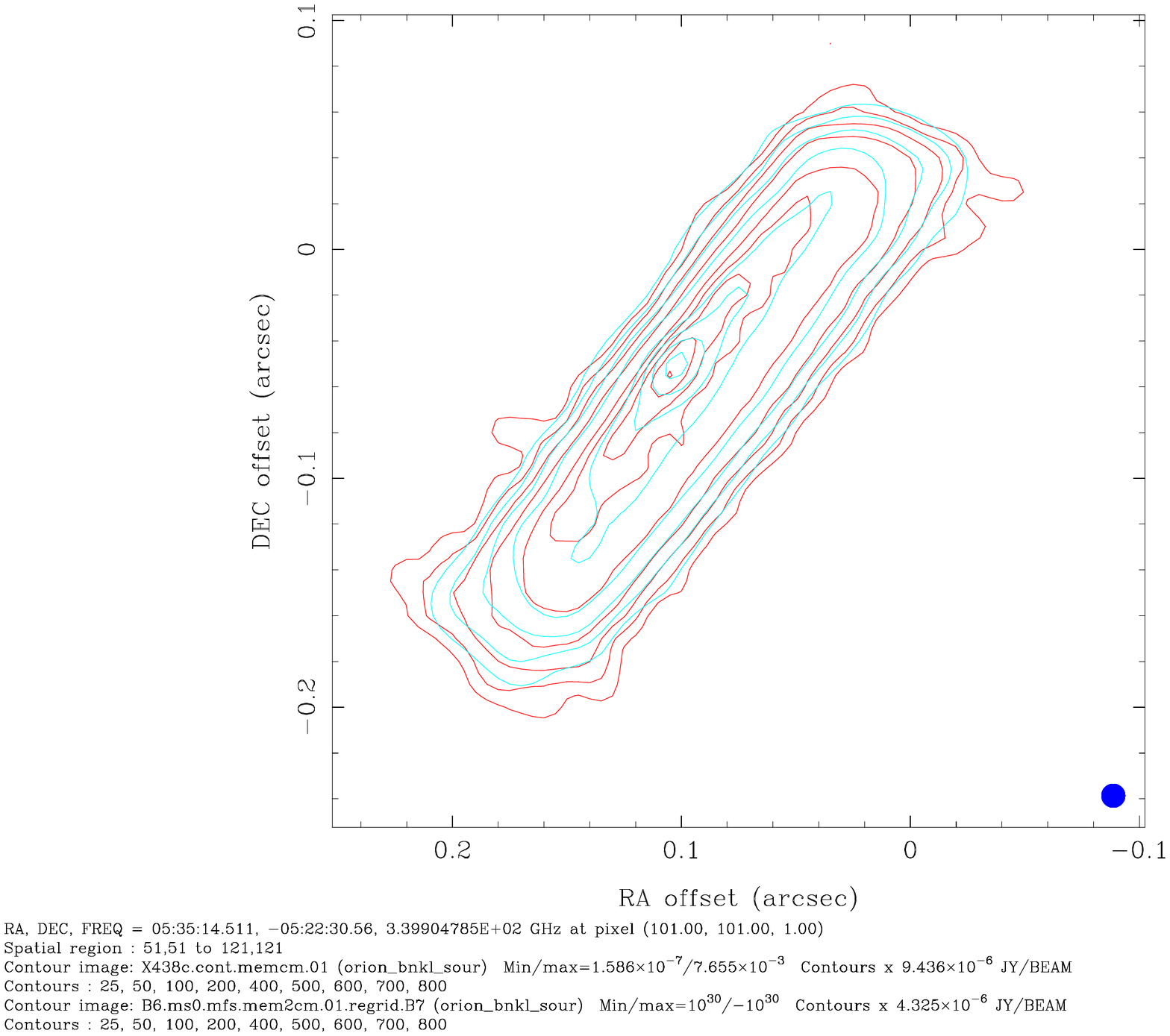}
\caption{Comparison of the Maximum Entropy images at 340 GHz and 223 GHz convolved with a 10 mas beam shown in lower right. 
 Contours are at 25, 50, 100, 200, 300, 400, 500, 600, 700, and 800 K.
 \label{fig:B7-B6.memcm01}}
\end{figure}

In Figure~\ref{fig:B7-B6.memcm01}, we compare Maximum Entropy images of the 340 GHz and 223 GHz continuum emission, both convolved by a 10 mas Gaussian beam.  As expected, the convolution
reduces the gradients at the edges of the disk, which are still steeper on the NE edge than
the SW edge.  In both images, the { brightness temperature} plateau of the disk,
ignoring the compact source on the NE side, is $\sim$400 K. The brightness falls to $\sim$200 K on the 223 GHz image, where the brightness on the 340 GHz image is still $\sim$400 K. If we attribute this to dust emission, then the dust opacity at 223 GHz is $\sim$0.7, where the emission is still optically thick at 340 GHz. At the disk edges, the logarithmic contour levels are
approximately evenly spaced, suggesting an exponential gradient in the emission.
The 1/$e$ scale height estimated from
the gradient between 400 and 25 K at the edge of the disk at 340 GHz $\sim$ 2 AU on the facing, NE side, to $\sim$ 4 AU on the SW side. 
{ We may be seeing deeper into the disk on the facing, NE side where the continuum emission is brighter.}
 

Since
the disk surface layers are being torn up by the outflow as the dust grains are destroyed, however, it is likely that the spectral index of the dust emission varies through the surface layers, and we do not make  further interpretation of the density profile of the disk surface. 
 
 \subsection{Salt Emission}
\begin{figure}
\includegraphics[width=1.0\columnwidth, clip, trim=3cm 3.7cm 2.5cm 1cm]{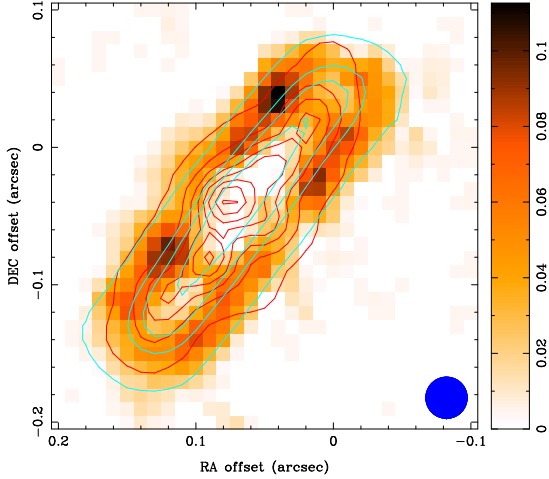}
\caption{NaCl emission at 232.51 GHz integrated over an LSR velocity range (-30 to 30)~\kms.
Red  contour levels at 99 GHz: 25 200 400 600 800 1000 1200 1400 K.
Blue contour levels at 340 GHz: 25 200 300 400 500 600 K. The color image shows the NaCl distribution. The 30 mas convolving beam FWHM is indicated in blue in the lower right.
\label{fig:B6-B3-NaCl}}
\end{figure}

\begin{figure}
\includegraphics[width=1.0\columnwidth, clip, trim=0.5cm 3.9cm 0.5cm 1cm]{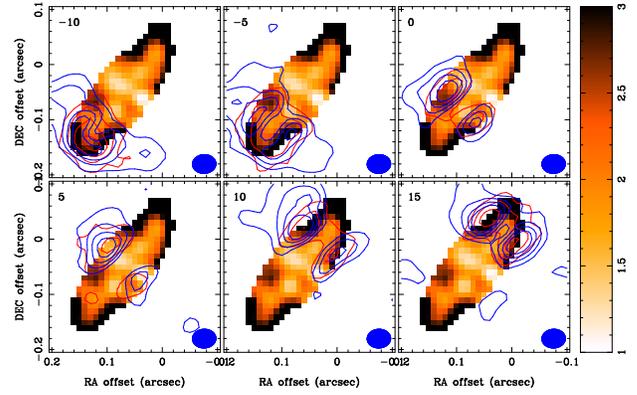}
\caption{Comparison of NaCl emission at 232.51 GHz and 335.51 GHz averaged in 5~\kms~channels, overlayed on the spectral index image.
Red  contours in NaCl line at 232.51 GHz: contour interval 20 K.
Blue contours in NaCl line at 335 GHz: contour interval 20 K.
The channel at 15~\kms is only 4.3~\kms wide to avoid contamination from adjacent spectral lines.
 The 30 $\times$ 40 mas convolving beam FWHM is 
 indicated in blue in the lower right.
\label{fig:B6-B3-NaCl-5kms}}
\end{figure}

Figure~\ref{fig:B6-B3-NaCl}  shows NaCl emission at 232.51 GHz overlayed on contours of 99 GHz and 340 GHz continuum images.  Salt emission is found in the dust layer
 at the surface of SrcI where there is a large spectral index gradient. A gradient in the
 dust opacity may help to explain the discrepancy in the NaCl excitation temperatures estimated from rotationally and vibrationally excited levels \citep{Ginsburg2019}. 
Figure~\ref{fig:B6-B3-NaCl-5kms} shows a comparison of NaCl emission at 232.51 GHz and 335.51 GHz averaged in 5~\kms~channels, overlaid on the spectral index image between 99 and 223 GHz. Both NaCl lines have peaks in the surface layers on both sides of the disk at close to the same positions. The 335 GHz NaCl $v=1$, $J=26-25$ line is strongly attenuated where the spectral index is less than $\sim$2. The 232 GHz NaCl $v=1$, $J=18-17$ line is less attenuated and shows a bridge of emission between the two peaks.
{ The NaCl emission is brighter on the NE side of the disk which is facing towards us. This could also be because there is less attenuation by the disk }.
Thus, we suggest that the
rotational temperatures, over a large range of observing frequencies, could be underestimated because of greater dust opacity at high frequencies,
 whereas the vibrational temperatures, over a small range of frequencies, are less affected.
 
\section{THE SRC I OUTFLOW}

\subsection{H$_2$O}
\begin{figure}
\includegraphics[width=1.0\columnwidth, clip, trim=3cm 3.7cm 2.7cm 1cm]{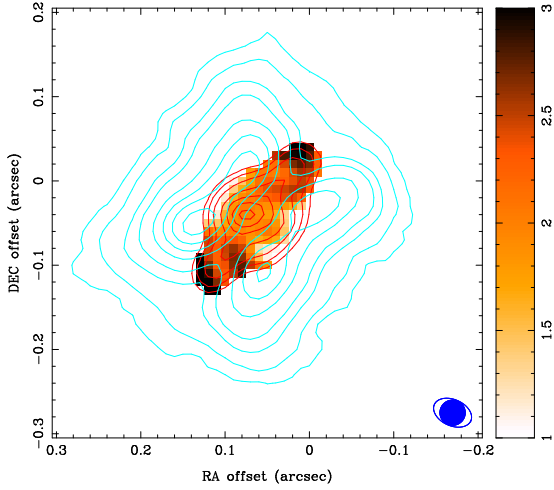}
\caption{H$_2$O emission at 232.687 GHz. The color image shows the 43 to 99 GHz spectral index
Red contours show the 43 GHz continuum emission at levels: 50, 100, 200, 400, 600, 800, 1000, 1200, and 1400 K.
Blue contours show the \ce{H2O} emission at 232.6867 GHz integrated over a velocity range (-50 to 50) km/s.
The contour levels are 0.05, 0.1, 0.2, 0.3, 0.5, 0.7, 0.9, and 1.1 $\times$  15345 K km s$^{-1}$.
 The 30 mas convolving beam FWHM for the continuum image and spectral index image is indicated in blue filled circle. The FWHM beam for the \ce{H2O} emission was 0.05 $\times$ 0.03 arcsec in PA 65$^{\circ}$ shown as the open ellipse in the lower right.
\label{fig:B3-B2.SI-H2O}}
\end{figure}

Figure~\ref{fig:B3-B2.SI-H2O} shows \ce{H2O} emission at 232.687 GHz overlayed on 43 GHz continuum, and spectral index images.
 H$_2$O emission extends from the surface of the SrcI disk into the outflow, and rotates with the SrcI disk. In contrast with
 the salt emission, the H$_2$O emission is more closely associated with the inner part of the disk, shown here at 43~GHz.
 { The H$_2$O emission may be in part excited by radiation from the central source, and is brighter on the NE side where we see deeper into the disk.}
 \citet{Hirota2017} present a model where the H$_2$O comes from the disk surface, and is driven by a magnetocentrifugal disk wind
 \citep{Matthews2010, Vaidya2013, Greenhill2013, Hirota2017}.

\subsection{AlO}
\begin{figure}
\includegraphics[width=1.0\columnwidth, clip, trim=3cm 4.4cm 3.2cm 1cm]{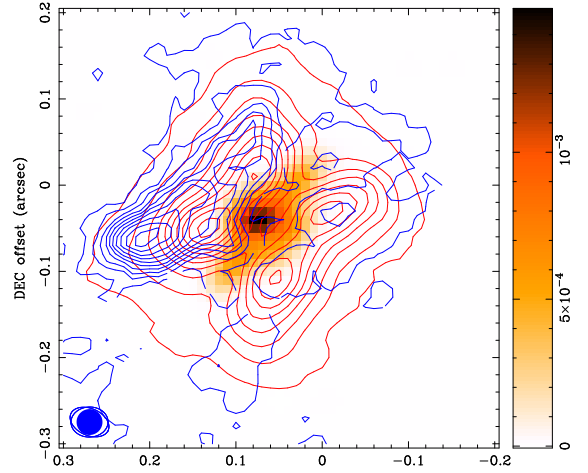}
\caption{AlO N=6-5 emission at 229.6938725 GHz
 overlayed on 43 GHz continuum, and \ce{H2O} emission at 232.6867 GHz. 
 Red contours show the \ce{H2O} emission at 232.6867 GHz integrated over a velocity range (-50 to 50) km/s. Lowest contour 933, contour interval 1866 K km s$^{-1}$.
Blue contours show the AlO N=6-5 emission integrated over -50 to 50~\kms. Lowest contour 225, contour interval 450 K km s$^{-1}$.  The 30 mas convolving beam FWHM for the 43 GHz continuum image is indicated in blue filled circle. The FWHM beams for the \ce{H2O} emission of 0.05 $\times$ 0.03 arcsec in PA 65$^{\circ}$, and for the AlO N=6-5, 0.04 $\times$ 0.03 arcsec in PA -88$^{\circ}$, are shown as the open ellipses in the lower left.
\label{fig:B3-B2.SI+H2O+AlO}}
\end{figure}

We mapped the AlO N=6-5 line at 229.69387 GHz, and 
the AlO N=9-8 line at 344.4537 GHz.
Figure~\ref{fig:B3-B2.SI+H2O+AlO} shows 229.694~GHz AlO emission overlayed on 43 GHz continuum and \ce{H2O} emission at 232.687 GHz.
AlO emission is coextensive with \ce{H2O}, but the peaks are further out, and brighter on the NE side of SrcI.
Since the NE side of the disk is facing towards us, we may be able to see deeper into the central regions.
{ The distributions of AlO emission and \ce{H2O} shown here are consistent with those presented by}
\citet{Tachibana2019} for AlO N=13-12 and N=17-16 emission lines at 497 and 650 GHz, and \ce{H2O} at 643 GHz \citep{Hirota2017}. \citet{Tachibana2019} attribute the distribution of AlO as due to its formation at the base of the outflow, and condensation further out into the
outflow.  
Our observations { with $\sim 4\times$ higher angular resolution,} suggest that AlO emission { peaks} downstream of the \ce{H2O}, and  may be produced by grain destruction and oxidized by O released by the dissociation of \ce{H2O}, or released as AlO further out in the outflow than the \ce{H2O} emission.

\begin{figure}
\includegraphics[width=1.0\columnwidth, clip, trim=3cm 4.4cm 3.2cm 1cm]{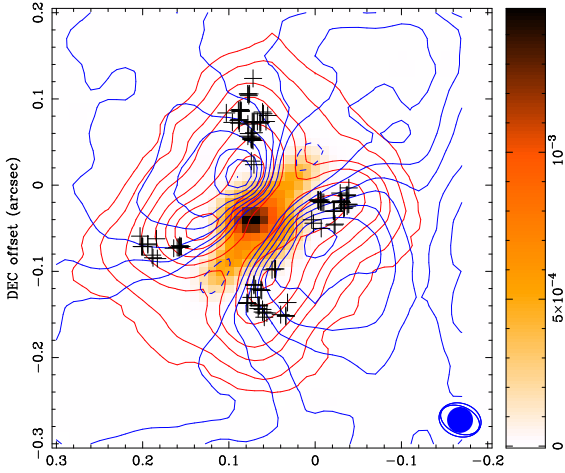}
\caption{SiO $v=0$ $J=5-4$  emission at 217.10498 GHz overlayed on 43 GHz continuum and \ce{H2O}
emission at 232.6867 GHz. The color image shows the 43GHz continuum emission
Red contours show the \ce{H2O} emission at 232.6867 GHz integrated over a velocity range of -50 to 50~\kms.
The contour levels are:
0.05, 0.1, 0.2, 0.3, 0.5, 0.7, 0.9, and 1.1 $\times$ 15345 K km s$^{-1}$.
Blue contours show the SiO $v=0$, $J=5-4$ emission integrated over the spectral line. Contour interval 2589 K km s$^{-1}$.
 The 30 mas convolving beam FWHM for the 43 GHz continuum image is indicated in blue filled circle. The FWHM beams for the \ce{H2O} emission of 0.05 $\times$ 0.03 arcsec in PA 65$^{\circ}$, and for the SiO $v=0$, $J=5-4$, 0.05 $\times$ 0.04 arcsec in PA 73$^{\circ}$, are shown as the open ellipses in the lower right. The black crosses indicate the centroid positions of the {\bf $v=1$, $J=2-1$ } SiO masers mapped
 by \citet{Issaoun2017} .
 \label{fig:B2+H2O+SiO54}}
\end{figure}


\subsection{SiO}
The SiO masers associated with SrcI have been mapped using VLBI at 43 GHz \citep{Kim2008, Matthews2010}, and 86 GHz \citep{Issaoun2017}. The $J=1-0$ and $J=2-1$,  $v=1$ { and $v=2$} masers positions lie in an X pattern, with blue-shifted emission in the south and east arms, and red-shifted emission in the north and west arms. The $J=1-0$ and $J=2-1$  $v=1$ { and $v=2$} masers can be interpreted as tracing a wide angle outflow arising from the surface of the almost edge on rotating disk \citep{Issaoun2017}.  No masers are located within a 14 AU band in PA $\sim$141$^{\circ}$~ that corresponds to the continuum emission from the disk \citep{Matthews2010}.
{ \citet{Goddi2009} modeled the SiO maser excitation including both radiative and collisional excitation. The model predicts densities $n_{H_2}$  = $10^8$ to $10^{10}$ cm$^{-3}$ for the $v=1$, and $10^9$ to $10^{11}$ cm$^{-3}$ for the $v=2$ transitions. The $v=2$ transition is favored where $T_k > 2000 K$, closer to SrcI than $v=1$ as is observed \citep{Kim2008, Matthews2010}
}.
Figure~\ref{fig:B2+H2O+SiO54} shows
SiO $v=0$, $J=5-4$  emission at 217.10498 GHz overlaid on the 43~GHz continuum and 232.69~GHz \ce{H2O} emission. The centroid positions of the $J=2-1$ masers mapped using the VLBA
are indicated by the black crosses which lie along the ridges of the peak \ce{H2O} emission. \ce{H2O} peaks closer to the disk, and SiO extends farther into the lobes.  The SiO $v=0$, $J=5-4$ distribution is similar to that of the SiO $v=2$, $J=10-9$ mapped by \citet{Kim2019}. 
\citet{Kim2019} also mapped the \ce{^29SiO} v=2 J=11-10 and SiO v=4 J=11-10 lines. The spatial and velocity distribution of these high excitation lines also suggest that they are associated with the base of the outflow at the surface of the disk. 

These distributions fit into a scenario where the SiO is formed from grain destruction following the dissociation of \ce{H2O}
\citep{Schilke1997}. In these models, grain mantles and grain cores are destroyed in shocks; Si is released into the gas phase, and then oxidized by dissociation products of \ce{H2O}. In the Src I disk
Si can be ablated from disrupted grains by shock velocities $>$25~\kms.
In the post shock gas, \ce{H2O} is depleted onto grains and converted to OH.
OH reacts with Si to form SiO, and \ce{SiO2} if OH is abundant. \citet{cohen2006}
note the absence of OH masers within 1300 AU of SrcI where SiO is prominent.
{ OH masers occur in densities $n_{H_2} \sim 10^7$ cm$^{-3}$, close to a shock front where the OH abundance is high. Gain path lengths 1 - 10 AU, not necessarily contiguous, suffice to produce OH masers \citep{Gray1992}, so either chemistry or maser excitation is responsible for their absence closer to SrcI. \citet{Greenhill2013} suggest that shocks from a magnetocentrifugal wind from SrcI, can inject energy further out in the
outflow where the outflow speed exceeds the Alfven and sound speeds}.
 SiO emission extends from the surface of SrcI into the extended outflow lobes, and rotates with SrcI.  
SiO is a good tracer of shocks. 
{ The implication is that the OH is depleted in the inner regions
where there is free Si. } Our data { suggest that chemistry and shock excitation play a significant role in determining these molecular distributions}.

\subsection{SiS}

\begin{figure*}
\includegraphics[width=2.0\columnwidth, clip, trim=0.5cm 0.5cm 0.5cm 0.5cm]{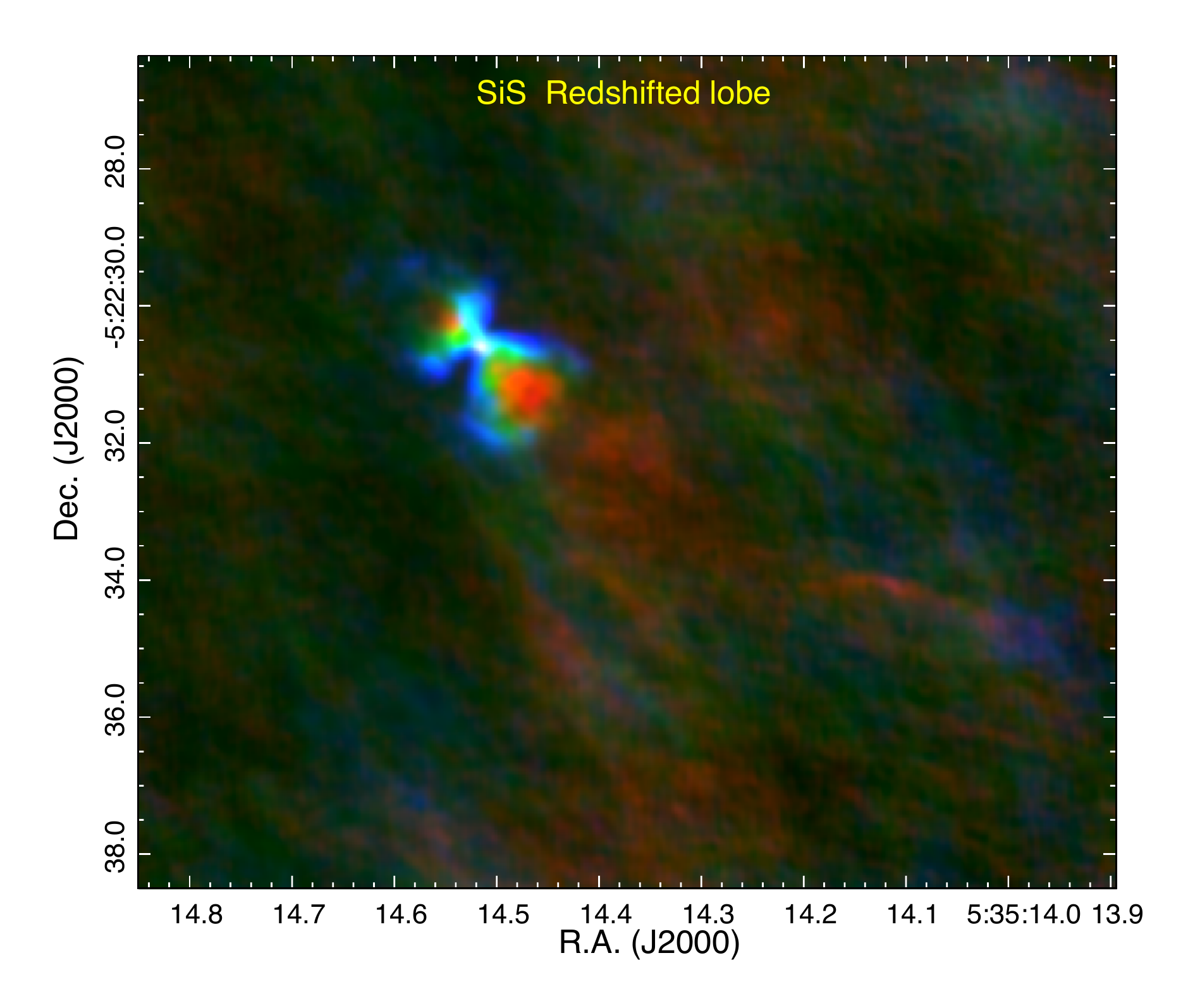}
\caption{Three-color image of SiS (217.81766 GHz) emission.
Blue +4 to +8 km/s,  Green +9 to +16 km/s,  Red +17 to +30 km/s.
{ The velocity ranges are chosen to highlight the enhancement of SiS emission around the SrcI outflow in the redshifted lobe.}
\label{fig:SiS_red}}
\end{figure*}

\begin{figure*}
\includegraphics[width=2.0\columnwidth, clip, trim=0.5cm 0.5cm 0.5cm 0.5cm]{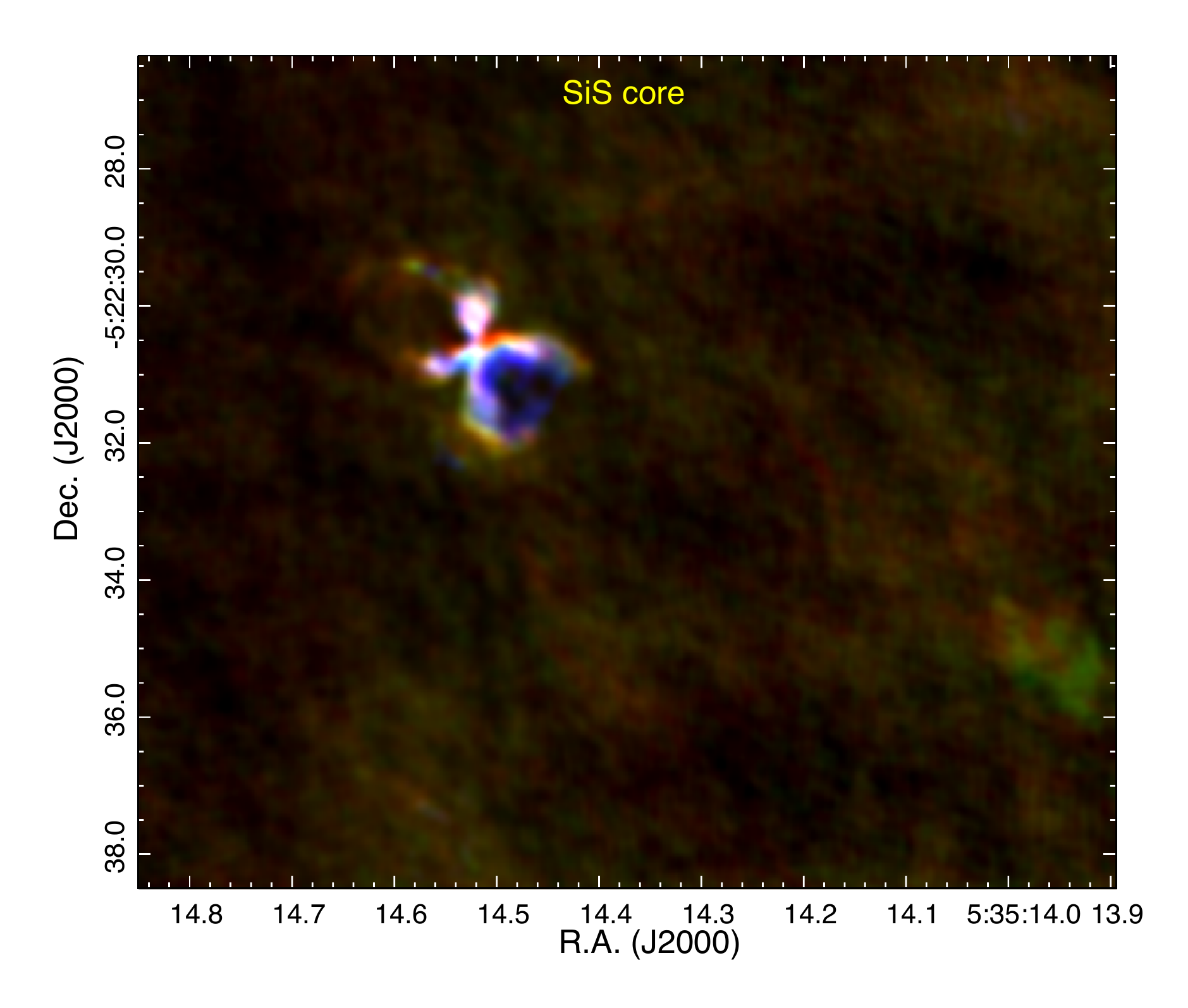}
\caption{Three-color image of SiS (217.81766 GHz) emission.
Blue -3 to -1 km/s,  Green 0 km/s,  Red +1 to +3 km/s.
{ The velocity ranges are chosen to highlight the enhancement of SiS emission around the SrcI outflow in the SiS core.}
\label{fig:SiS_core}}
\end{figure*}

\begin{figure*}
\includegraphics[width=2.0\columnwidth, clip, trim=0.5cm 0.5cm 0.5cm 0.5cm]{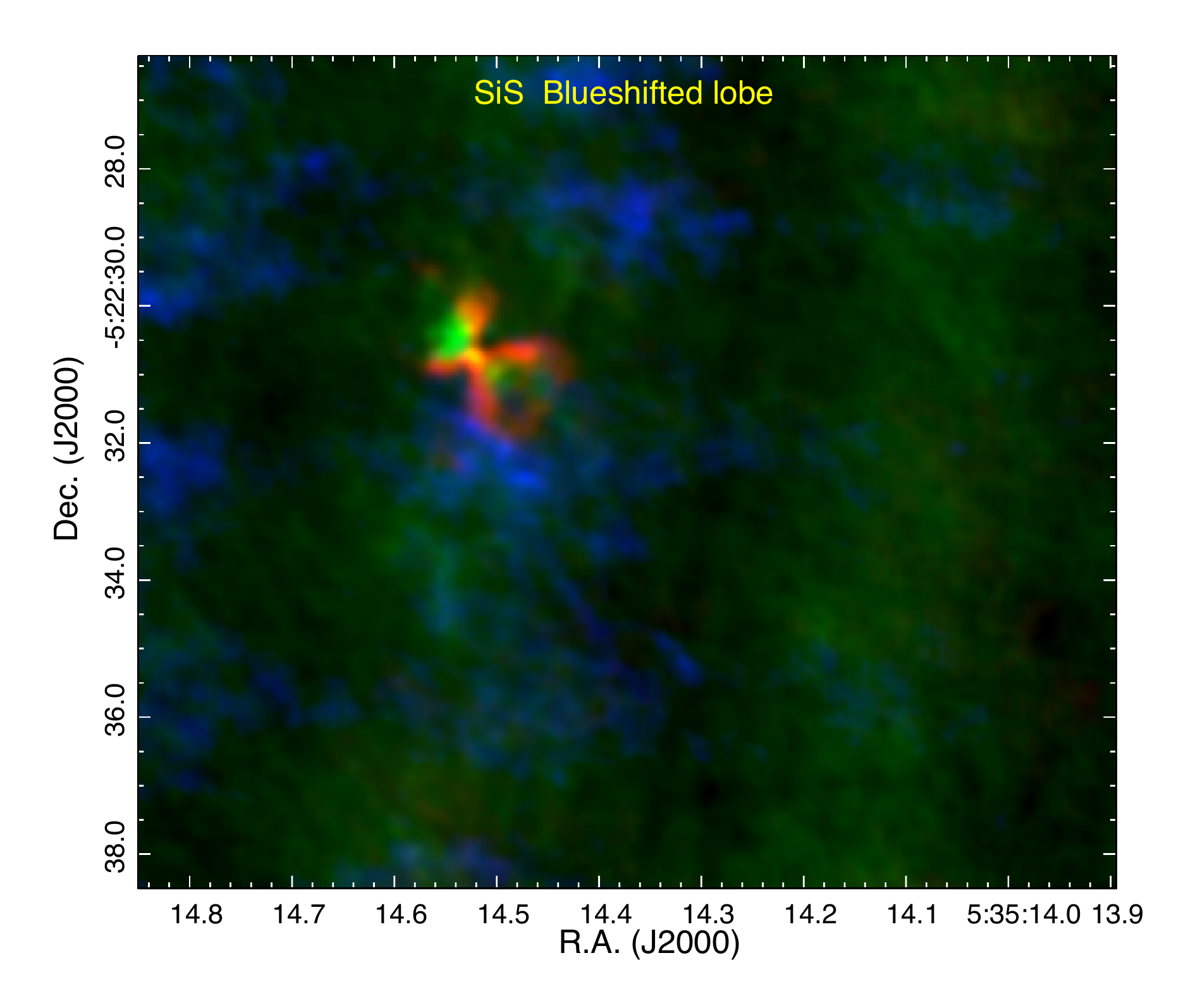}
\caption{Three-color image of SiS (217.81766 GHz) emission.
Blue -23 to -12 km/s,  Green -11 to -7 km/s,  Red -6 to -3 km/s.
{ The velocity ranges are chosen to highlight the enhancement of SiS emission around the SrcI outflow in the blueshifted lobe.}
\label{fig:SiS_blue}}
\end{figure*}

 Figure~\ref{fig:SiS_red}, Figure~\ref{fig:SiS_core},
 and Figure~\ref{fig:SiS_blue}, show three-color images of the 217.817~GHz SiS line. These color images clearly show the wide-angle bipolar wind from Src I.  This feature in SiS is highly limb-brightened.  In addition to motion away from Src I with the blueshifted lobe extending NE and redshifted lobe extending SW, the SiS outflow exhibits rapid expansion away from the outflow axis.  The SiS outflow is highly asymmetric with a stubby NE blue lobe and a much longer SW red lobe.  Weak emission can be traced to the lower-right edge of the figures above. Also note the patch of SIS emission near the SW corner of the image at the "core" velocities. 

\begin{figure*}
\includegraphics[width=2.0\columnwidth, clip, trim=3cm 4.3cm 2.7cm 1cm]{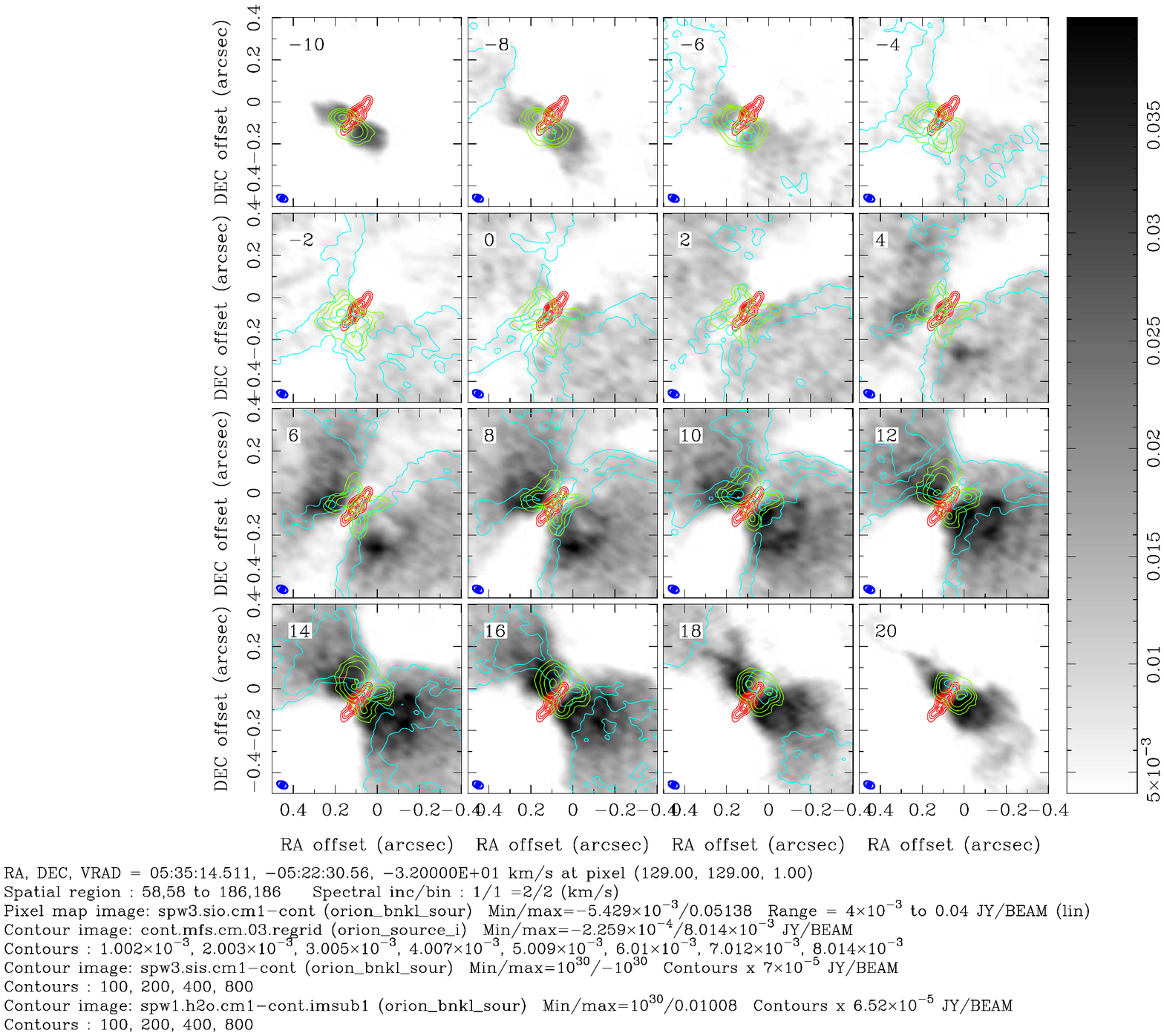}
\caption{Grey scale shows SiO (217.10498 GHz) in 2~\kms~channels and ranges from 4 to 40 mJy beam$^{-1}$.
Red contours show 99 GHz continuum with contour levels of 1, 2, 3, 4, 5, 6, 7, and 8 mJy beam$^{-1}$ in a 30 mas beam.
Blue contours show SiS (217.81766 GHz) in 2~\kms~channels, with contour levels at 100, 200, and 400 K in a 50 $\times$ 30 mas beam.
Green contours show \ce{H2O} (232.6867 GHz) in 2~\kms~channels, with contour levels at 100, 200, and 400 K in a 50 $\times$ 30 mas beam.
 The convolving beams is indicated in blue in the lower left.
\label{fig:sio+sis+H2O}}
\end{figure*}

Figure~\ref{fig:sio+sis+H2O} compares the 217.817~GHz SiS line with SiO and \ce{H2O} emission in 2~\kms~velocity channels.  Like SiO and \ce{H2O}, SiS shows rotation close to SrcI.  
Approximately 50 AU above and below the SrcI disk, close to the maximum extent of the \ce{H2O} emission region, the SiO column expands abruptly into a turbulent wide angle outflow.
Filaments of SiS emission are particularly
prominent along the edges of the outflow.

The distributions of SiO and SiS are consistent with the model of \citet{Podio2017}, who argue that SiO results from dust grain destruction, whereas SiS is the product of gas phase chemistry.  SiO and SiS are also seen in L1157-B1 in
the outflow from this low mass protostar \citep{Podio2017}. As in Orion, SiO and SiS have different spatial distributions: SiO is seen where the molecular jet impacts the outflow cavity wall, but SiS is detected only at the leading edge of the outflow. 
\citet{Zanchet2018} also considered silicon, oxygen, and sulfur chemistry in postshock gas in protostellar outflows. They found  that the main SiS-forming reactions are Si + SO and Si + \ce{SO2}, and that SiS is efficiently destroyed through reaction with atomic oxygen. 

\subsection{SO and \ce{SO2}}
\begin{figure*}
\includegraphics[width=2.0\columnwidth, clip, trim=3cm 3.0cm 2.5cm 1cm]{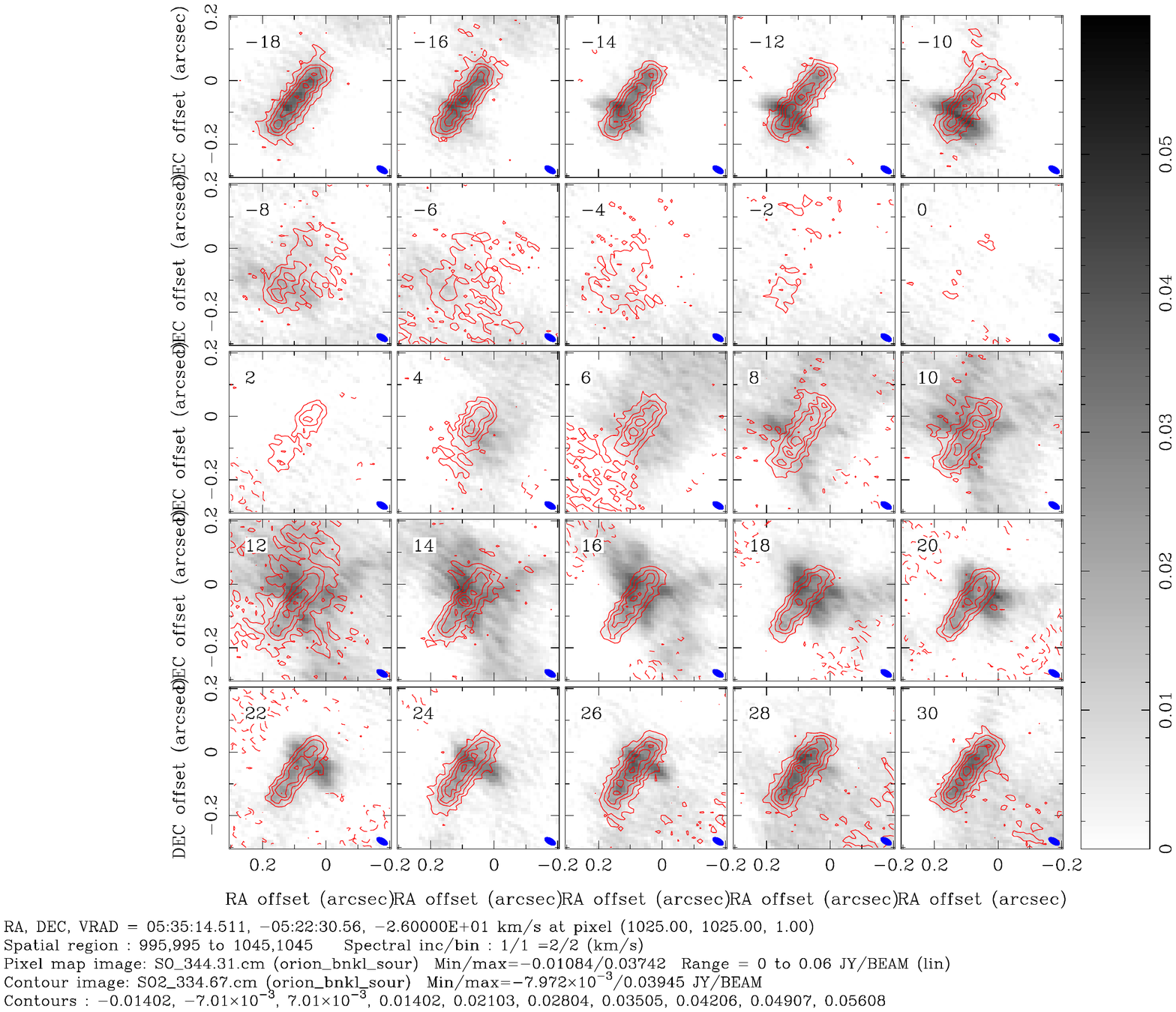}
\caption{SO and \ce{SO2} emission and absorption across SrcI. Both images
plot line emission + continuum emission from the disk to show the line
absorption across the disk. The grey scale image shows the SO at 344.31 GHz. Peak 0.039 Jy/beam = 655 K. The contours show the \ce{SO2} at 334.67 GHz.
Contour interval 129 K, peak 729 K. The convolving beam 36 $\times$ 17 mas FWHM in PA = 59 $\deg$ is indicated in blue in the lower right.
\label{fig:so-SO2}}
\end{figure*}

Figure~\ref{fig:so-SO2} shows the SO emission at 344.31 GHz, and  \ce{SO2} at 334.67 GHz in 2~\kms~velocity channels.
Both SO and \ce{SO2} are heavily absorbed across SrcI at velocities of -8 to +4~\kms.  
This is consistent with the 0.2\arcsec\ resolution spectra of SrcI plotted by \citet{Plambeck2016}, in which sulfur-bearing species, including CS and \ce{H2S} as well as SO and \ce{SO2}, exhibit prominent blueshifted absorption profiles.  Since the velocity width of the absorption is comparable with the halfwidth of the SiO emission lines, it is likely that the absorbing molecules are located in the (cooler) outer layers of the SrcI outflow, rather than in unrelated foreground gas.  Examining
Figure~\ref{fig:so-SO2}, one sees that SO emission (gray scale) follows the rotation of the outflow { close to the disk}, while \ce{SO2} { does not show this small scale structure}. Models of C-type shocks with velocities 5-40 km/s and densities $10^4 - 10^6$  cm$^{-3}$ can enhance SO and \ce{SO_2}  abundances by 2 orders of magnitude. SO decreases quickly { after} the passage of a shock, whilst \ce{SO_2} is enhanced during and for some time after the shock \citep{Pineau1993}.

The SO and \ce{SO2} distributions are over-resolved in these ALMA observations.  
At lower resolution,  SO and \ce{SO2} are seen as a shell of expanding
gas, with enhanced abundances where the outflow from SrcI impacts dense clumps around the edges of the outflow \citep{Wright1996,wright2017}. 
\citet{Goddi2011b} suggested that the hot core is excited by the SrcI outflow impacting a dense core.
The correspondence of these continuum clumps with the edge of the outflow is shown in Figure~\ref{fig:B7-SiO-Uline}.

\subsection{Unidentified line emission along outflow axis}
\begin{figure*}
\includegraphics[width=2.0\columnwidth, clip, trim=3cm 3.6cm 2.5cm 1cm]{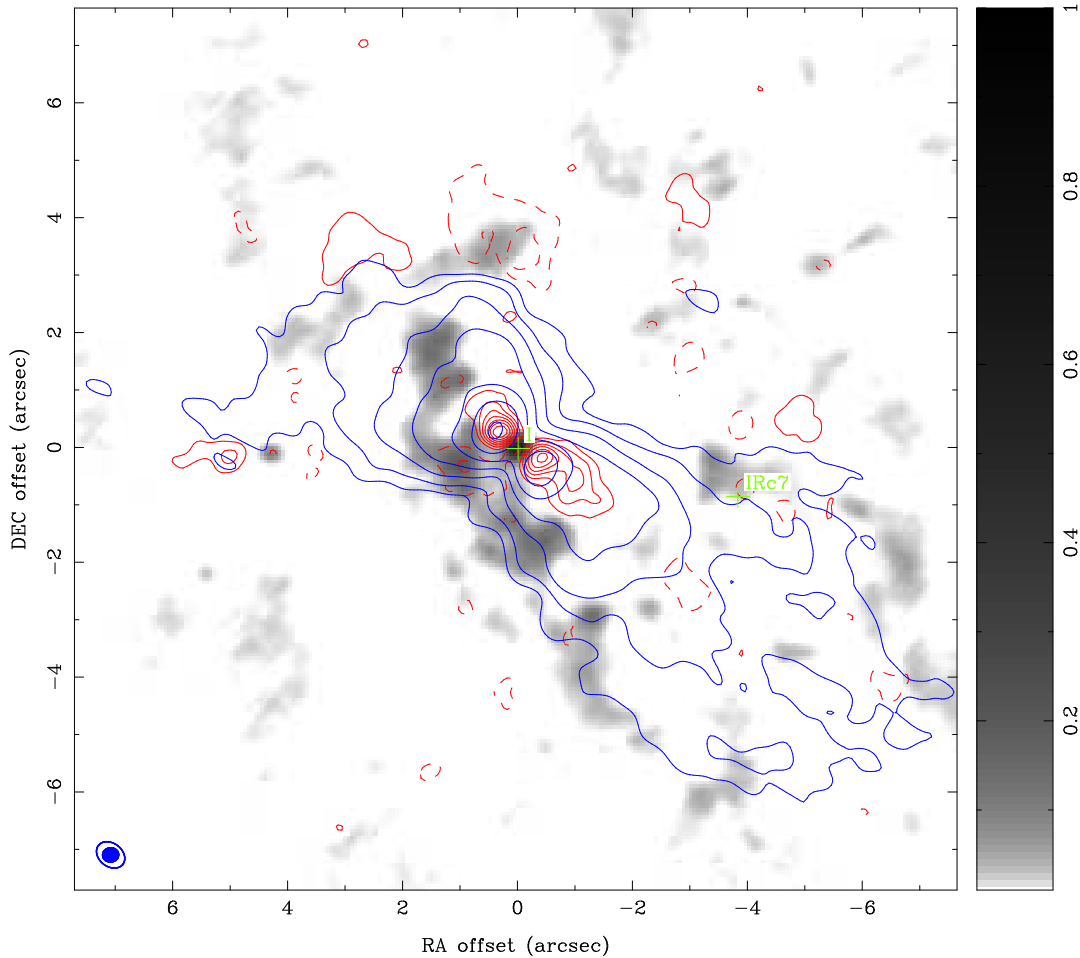}
\caption{Continuum emission around
the edges of the SiO J=2-1 v=0 line (Blue contour levels 0.05 0.1 0.2 0.4 0.8 1.6 3.2 Jy/beam (746 K/Jy), integrated over -10 to + 20 km/s). Red contours map the integrated emission over 11 km/s from an unidentified  line at
354.4945 GHz. Contour interval 14.8 K. The grey scale image shows the continuum emission at 348 GHz (141 K/Jy).  The convolving beams, 0.28 $\times$ 0.26 $''$ for the unidentified line and continuum emission, and 0.54 $\times$ 0.41 $''$ for the SiO are indicated in blue in the lower left.
\label{fig:B7-SiO-Uline}}
\end{figure*}

Figure~\ref{fig:B7-SiO-Uline} 
also
shows an unidentified line that appears in two lobes along the outflow axis, centered on SrcI.
Assuming the line is centered at 5\kms, like the SiO, the rest frequency of the U-line
is 354.4945 GHz.
Gaussian fits to the spectra at the 2 peaks give velocities 
4.3 +/- 0.1,  FWHM 5.8 +/- 0.2 km/s for the NE peak, and
+5.7 +/- 0.1 km/s, FWHM 6.7 +/- 0.2 km/s for the SW peak.

The U-line is blueshifted to NE like the SiO outflow, but the FWHM  is smaller, and the emission is more confined along the outflow axis. 
There is a vibrationally excited HCN line at 354.4604 GHz $\sim$26km/s from the U-line centered on 354.4945 GHz in this image.  The U-line image has well defined spectral peaks, and has low sidelobes and is probably therefore quite compact, whereas
the image of the HCN line emission has large sidelobes suggesting missing larger scale structure.
The U-line could be an SiO isotopologue or a high vibrational transition illuminated by the central source.
 The Lovas catalog lists an unidentified line at 354.4968~GHz. In the CDMS catalog 
 there are many lines from 354.49 to 354.50 GHz. Most of them are organic molecules except TiO$_2$ at 354.4977~GHz, but it is also not convincing.



\begin{deluxetable*}{cccccc}
\tabletypesize{\small}
\tablecaption{Observations}
\tablecolumns{6}
\tablenum{1}
\tablehead{
\colhead{freq} &
\colhead{project code} &
\colhead{date} &
\colhead{time} &
\colhead{synth beam}  &
\colhead{baseline} \\
\colhead{(GHz)} & & & 
\colhead{(min)} &
\colhead{(milliarcsec)} &
\colhead{(meters)}
}
\startdata
 43  &  VLA/18A-136     &  2018-03-06  &  291  &   56$\times$44 at PA 33  & 500 - 36600  \\
 99  &  2017.1.00497.S	&  2017-10-12  &  158  &   45$\times$36 at PA 47  & 40 - 16200 \\
 216-220  & 2013.1.005446.S  & 2014-12 to 2015-04  &   15    &   1500$\times$930 at PA -8 & 14 - 330 \\
224  &  2016.1.00165.S	&  2017-09-19  &   44  &   39$\times$19 at PA 66  & 40 - 10500 \\
340  &	2016.1.00165.S  &  2017-11-08  &   45  &   26$\times$11 at PA 58  & 90 - 12900 \\
350  &	2012.1.00123.S  &  2014-07-26  &   24  &   276$\times$260 at PA 85 & 30 - 730 \\
\enddata
\end{deluxetable*}

\begin{deluxetable*}{CCCC}   
\tabletypesize{\small}
\tablecaption{Measured sizes and flux densities for source I}
\tablecolumns{4}
\tablenum{2}
\tablehead{
\colhead{freq} &
\colhead{beam} &
\colhead{deconvolved size} &
\colhead{integrated flux} \\
\colhead{(GHz)} &
\colhead{(arcsec, PA)} &
\colhead{(arcsec, PA)} &
\colhead{(mJy)}
}
\startdata
%
43 &
0.06 \times 0.04, 33\degr &
0.099 \pm 0.002 \times 0.057 \pm 0.002, -40.4\degr \pm 2.7\degr &
10 \pm 1 \\
%
99 &
0.04 \times 0.04, 47\degr &
0.151 \pm 0.005 \times 0.044 \pm 0.002, -37.8\degr \pm 1.3\degr &
58 \pm 6 \\
%
224 &
0.04 \times 0.02, 66\degr &
0.197 \pm 0.003 \times 0.042 \pm 0.003, -37.3\degr \pm 0.4\degr &
256 \pm 25\\
%
340 &
0.03 \times 0.01, 58\degr &
0.234 \pm 0.005 \times 0.042 \pm 0.002, -37.4\degr \pm 0.3\degr &
630 \pm 63 \\
\enddata
\end{deluxetable*}
 
\section{DISCUSSION and CONCLUSIONS}

{ Our analysis has focused on understanding whether the chemistry seen in SrcI is a result of an ``anomalous" history of interactions with its environment, or whether it is directly attributable to the physical conditions of SrcI and its associated outflow, making SrcI a paradigm for the study of high-mass star formation.}

The distributions of \ce{H2O}, SiO, AlO, and SiS lend strong support to a model in which dust grains are ablated
and destroyed close to the disk surface, producing an oxygen rich outflow.
The strong SiO maser emission, and AlO  mapped in the
outflow close to the disk, suggest that refractory grain cores as well as the grain mantles are destroyed. \citet{Lenzuni1995} investigated the evaporation of dust grains in protostellar cores. Carbon grains are destroyed at temperatures $\sim$ 800 -- 1150 K. Silicate grains are evaporated at temperatures $\sim 1300$ K, followed by AlO at $\sim$ 1700 K. SiO and AlO may be released directly from the grains, or may be formed in the gas phase by the oxidation of Si and Al.
{ Thermal emission from the ground state, $v=0$,} SiO traces a turbulent, wide angle outflow extending over 1000 AU
from a position close to the maximum extent of the \ce{H2O} emission from the disk \citep{Plambeck2009}.
SiS traces a more filamentary structure which is prominent at edges of extended SiO outflow.
\citet{Zanchet2018}, in a study of silicon, oxygen, and sulfur chemistry, found that the main formation for SiS are from SO and \ce{SO2} which are seen as a shell around the edges of the outflow \citep{Wright1996}.

 \begin{figure*}
\includegraphics[width=2.0\columnwidth, clip, trim=3cm 3.6cm 2.5cm 1cm]{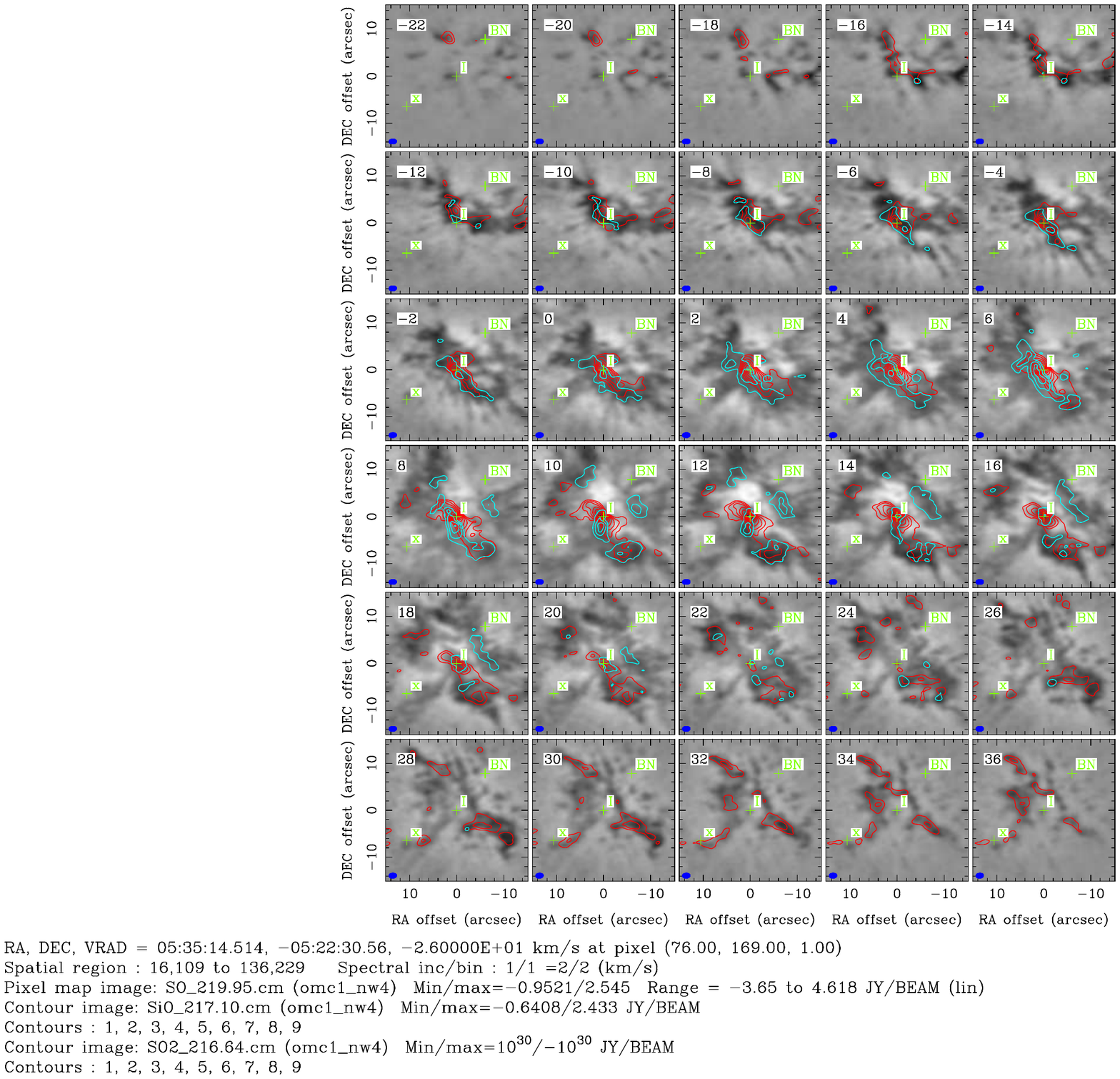}
\caption{SiO, SO and \ce{SO2} emission associated with the explosion resulting from the SrcI/BN interaction. The grey scale image shows the SO at 219.95 GHz. The extended SO emission is
heavily resolved in these ALMA observations, but nicely shows
the filamentary structure.
The gray scale range is -65 to +85 K. The red contours show the SiO emission at 217.10 GHz, and the blue contours show the \ce{SO2} emission at 216.64 GHz. The contour interval is 19 K in both SiO and \ce{SO2}. The convolving beam FWHM of 1.5\arcsec~$\times$ 0.9\arcsec~is indicated in blue in the lower left.
\label{fig:sio+so+SO2}}
\end{figure*}

One might ask if the chemistry of the bipolar outflow is distinct from that of the larger ``finger system'' produced by the BN-SrcI explosive event.  The finger system is over an arcminute in extent, and so is almost completely resolved out in our high resolution images.  
CO emission from the fingers was mapped by \citet{Bally2017} with approximately 1\arcsec\ resolution using mosaic observations with ALMA.  Those data also cover transitions of SiO (J=5-4 v=0; 217.105~GHz), SO (219.95~GHz), and \ce{SO2} (216.64 GHz).  We used the mosaic ALMA data to generate 2~\kms\ channel maps of these three molecular lines over a 30\arcsec\ region centered on SrcI.  These images are compared in Figure~\ref{fig:sio+so+SO2}.
Both the SiO and SO images exhibit streamers similar those seen in CO; the situation for \ce{SO2} is less clear.  
The SiO fingers east of SrcI at velocities 
$>$30~\kms\ were noted earlier by \citet{Plambeck2009} (see their Figure 4).
 It is extremely difficult to distinguish the outflow from the finger system in any of these images.
 In these molecular lines, at least, there appear to be no clear chemical or excitation differences between the bipolar outflow and the explosive event.
 
 It is possible that the SrcI/BN interaction, resulting in a collapse of a binary protostar in SrcI, and the disruption of the disk around SrcI, dredged up the inner part of the disk, and made oxygen and sulfur rich material available for both the SrcI outflow and the explosion, and moved dust grains containing refractory material like AlO and salts to the outer parts of the disk where they could be ablated by the SrcI outflow. The energy for the ejection of SrcI and BN, and the explosion in the gas must have come from gravitational binding energy exceeding 10$^{48}$ ergs \citep{Bally2017}.
The violence of the interaction $\sim$550 years ago would have strongly shocked the disk, resulting in chemistry. The AlO and salts could be a bi-product of this interaction.   As the disk relaxed from its perturbed state, it could have triggered a major accretion event onto the central binary or a merger { which produced } the current Src I outflow { with}
 about the same age. At 50 AU, the disk is well inside the Src I gravitational radius for a 10 km s$^{-1}$ ejection speed. 
Thus the disk is likely made of material that was already bound to one of more stars involved in the interaction - no need for Bondi-Hoyle accretion over the last 500 years. \citet{Moeckel2012} studied the chances of disk survival in binary-single stellar interactions with N-body simulations.
 
 Observations of other high mass protostar outflows are required to see if \ce{H2O}, sulfur and Si
 molecules, and salts are common or if the SrcI outflow is unusual.
  We adopt as a working hypothesis, that the rich chemistry seen in SrcI is a direct consequence of the outflow from a high mass protostar and does not depend on the particular environment and history of SrcI.

If, indeed, Orion SrcI, as the closest and best studied example of a young high mass
protostar, is a paradigm model, then future observations of other high-mass star formation regions will have many tools available. The resolution and sensitivity of our ALMA
observations are sufficient to detect disks and outflows similar to those seen in SrcI
out to $\sim$2~kpc. Our $\sim$50 mas resolution images (100 AU at 2 kpc) have an RMS brightness sensitivity $\sim$10 K at 2\kms~velocity resolution which would be sufficient to detect and image SiO, SiS, \ce{H2O}, and salt distributions in outflows from putative disks around other high mass protostars.

Our observations from 43 to 350 GHz allow us to map the dust opacity at the surface
of the disk around SrcI.
NaCl and KCl trace the rotation of the surface layers of SrcI, allowing us to estimate
the mass of the protostar.
Salt emission is confined to the dust layer where there is a large spectral index gradient.
NaCl excitation temperatures estimated from rotational transitions over a large range of  frequencies could be underestimated because of greater dust opacity at high frequencies,
whereas the vibrational temperatures, estimated over a small range of frequencies, are less affected.

\ce{H2O} emission maps the distribution of material released from grain mantles.
SiO maps the more extended outflow. 
Maps of SiS and SO in other outflows could support the role of gas phase chemistry
in the post shock gas in the outflow. 

\acknowledgments
This paper makes use of the following ALMA data:
ADS/JAO.ALMA\#2012.1.00123.S, ADS/JAO.ALMA\#2013.1.00546.S, ADS/JAO.ALMA\#2016.1.00165.S, ADS/JAO.ALMA\#217.1.00497.S. 
ALMA is a partnership of ESO (representing its member states), NSF (USA) and NINS (Japan), together with NRC(Canada) and NSC and ASIAA (Taiwan), in cooperation with the Republic of
 Chile. The Joint ALMA Observatory is operated by ESO, AUI/NRAO and NAOJ.

The National Radio Astronomy Observatory is a facility of the National Science Foundation operated under cooperative agreement by Associated Universities, Inc.
TH is financially supported by the MEXT/JSPS KAKENHI Grant
Numbers 24684011, 17K05398 and 18H05222.
We thank the referee, Paul Ho, for a careful reading which has improved the presentation of this paper.

%

\vspace{5mm}
\facilities{ALMA, JVLA}


\software{Miriad \citep{Sault1995}}

\bibliographystyle{aasjournal}
\bibliography{SrcI.bib}

\begin{thebibliography}{}
\expandafter\ifx\csname natexlab\endcsname\relax\def\natexlab#1{#1}\fi
\providecommand{\url}[1]{\href{#1}{#1}}

\bibitem[{{Bally} {et~al.}(2017){Bally}, {Ginsburg}, {Arce}, {Eisner},
  {Youngblood}, {Zapata}, \& {Zinnecker}}]{Bally2017}
{Bally}, J., {Ginsburg}, A., {Arce}, H., {et~al.} 2017, \apj, 837, 60

\bibitem[{{Cohen} {et~al.}(2006){Cohen}, {Gasiprong}, {Meaburn}, \&
  {Graham}}]{cohen2006}
{Cohen}, R.~J., {Gasiprong}, N., {Meaburn}, J., \& {Graham}, M.~F. 2006,
  \mnras, 367, 541

\bibitem[{{Ginsburg} {et~al.}(2018){Ginsburg}, {Bally}, {Goddi}, {Plambeck}, \&
  {Wright}}]{Ginsburg2018}
{Ginsburg}, A., {Bally}, J., {Goddi}, C., {Plambeck}, R., \& {Wright}, M. 2018,
  \apj, 860, 119

\bibitem[{Ginsburg {et~al.}(2019)Ginsburg, McGuire, Plambeck, Bally, Goddi, \&
  Wright}]{Ginsburg2019}
Ginsburg, A., McGuire, B., Plambeck, R., {et~al.} 2019, The Astrophysical
  Journal, 872, 54.
\newblock \url{https://doi.org/10.3847%2F1538-4357%2Faafb71}

\bibitem[{{Goddi} {et~al.}(2009){Goddi}, {Greenhill}, {Chandler}, {Humphreys},
  {Matthews}, \& {Gray}}]{Goddi2009}
{Goddi}, C., {Greenhill}, L.~J., {Chandler}, C.~J., {et~al.} 2009, \apj, 698,
  1165

\bibitem[{{Goddi} {et~al.}(2011{\natexlab{a}}){Goddi}, {Greenhill},
  {Humphreys}, {Chandler}, \& {Matthews}}]{Goddi2011b}
{Goddi}, C., {Greenhill}, L.~J., {Humphreys}, E.~M.~L., {Chandler}, C.~J., \&
  {Matthews}, L.~D. 2011{\natexlab{a}}, \apjl, 739, L13

\bibitem[{{Goddi} {et~al.}(2011{\natexlab{b}}){Goddi}, {Humphreys},
  {Greenhill}, {Chandler}, \& {Matthews}}]{Goddi2011}
{Goddi}, C., {Humphreys}, E.~M.~L., {Greenhill}, L.~J., {Chandler}, C.~J., \&
  {Matthews}, L.~D. 2011{\natexlab{b}}, \apj, 728, 15

\bibitem[{{G{\'o}mez} {et~al.}(2008){G{\'o}mez}, {Rodr{\'{\i}}guez}, {Loinard},
  {Lizano}, {Allen}, {Poveda}, \& {Menten}}]{Gomez2008}
{G{\'o}mez}, L., {Rodr{\'{\i}}guez}, L.~F., {Loinard}, L., {et~al.} 2008, \apj,
  685, 333

\bibitem[{{Gray} {et~al.}(1992){Gray}, {Field}, \& {Doel}}]{Gray1992}
{Gray}, M.~D., {Field}, D., \& {Doel}, R.~C. 1992, \aap, 262, 555

\bibitem[{{Greenhill} {et~al.}(2013){Greenhill}, {Goddi}, {Chandler},
  {Matthews}, \& {Humphreys}}]{Greenhill2013}
{Greenhill}, L.~J., {Goddi}, C., {Chandler}, C.~J., {Matthews}, L.~D., \&
  {Humphreys}, E.~M.~L. 2013, \apjl, 770, L32

\bibitem[{{Hirota} {et~al.}(2014){Hirota}, {Kim}, {Kurono}, \&
  {Honma}}]{Hirota2014}
{Hirota}, T., {Kim}, M.~K., {Kurono}, Y., \& {Honma}, M. 2014, \apjl, 782, L28

\bibitem[{{Hirota} {et~al.}(2017){Hirota}, {Machida}, {Matsushita}, {Motogi},
  {Matsumoto}, {Kim}, {Burns}, \& {Honma}}]{Hirota2017}
{Hirota}, T., {Machida}, M.~N., {Matsushita}, Y., {et~al.} 2017, Nature
  Astronomy, 1, 0146

\bibitem[{{Issaoun} {et~al.}(2017){Issaoun}, {Goddi}, {Matthews}, {Greenhill},
  {Gray}, {Humphreys}, {Chand ler}, {Krumholz}, \& {Falcke}}]{Issaoun2017}
{Issaoun}, S., {Goddi}, C., {Matthews}, L.~D., {et~al.} 2017, \aap, 606, A126

\bibitem[{{Kim} {et~al.}(2019){Kim}, {Hirota}, {Machida}, {Matsushita},
  {Motogi}, {Matsumoto}, \& {Honma}}]{Kim2019}
{Kim}, M.~K., {Hirota}, T., {Machida}, M.~N., {et~al.} 2019, \apj, 872, 64

\bibitem[{{Kim} {et~al.}(2008){Kim}, {Hirota}, {Honma}, {Kobayashi},
  {Bushimata}, {Choi}, {Imai}, {Iwadate}, {Jike}, {Kameno}, {Kameya},
  {Kamohara}, {Kan-Ya}, {Kawaguchi}, {Kuji}, {Kurayama}, {Manabe}, {Matsui},
  {Matsumoto}, {Miyaji}, {Nagayama}, {Nakagawa}, {Oh}, {Omodaka}, {Oyama},
  {Sakai}, {Sasao}, {Sato}, {Sato}, {Shibata}, {Tamura}, \&
  {Yamashita}}]{Kim2008}
{Kim}, M.~K., {Hirota}, T., {Honma}, M., {et~al.} 2008, \pasj, 60, 991

\bibitem[{{Kounkel} {et~al.}(2018){Kounkel}, {Covey}, {Su{\'a}rez},
  {Hernandez}, {Stassun}, {Jaehnig}, {Feigelson}, {Pe{\~n}a Ram{\'\i}rez},
  {Roman-Lopes}, {Da Rio}, {Stringfellow}, {Kim}, {Borissova},
  {Fern{\'a}ndez-Trincado}, {Burgasser}, {Garc{\'\i}a-Hern{\'a}ndez}, {Zamora},
  {Pan}, \& {Nitschelm}}]{Kounkel2018}
{Kounkel}, M., {Covey}, K., {Su{\'a}rez}, Genaro a
  nd~{Rom{\'a}n-Z{\'u}{\~n}iga}, C., {et~al.} 2018, \aj, 156, 84

\bibitem[{{Lenzuni} {et~al.}(1995){Lenzuni}, {Gail}, \&
  {Henning}}]{Lenzuni1995}
{Lenzuni}, P., {Gail}, H.-P., \& {Henning}, T. 1995, \apj, 447, 848

\bibitem[{{Matthews} {et~al.}(2010){Matthews}, {Greenhill}, {Goddi}, {Chand
  ler}, {Humphreys}, \& {Kunz}}]{Matthews2010}
{Matthews}, L.~D., {Greenhill}, L.~J., {Goddi}, C., {et~al.} 2010, \apj, 708,
  80

\bibitem[{{Menten} {et~al.}(2007){Menten}, {Reid}, {Forbrich}, \&
  {Brunthaler}}]{Menten2007}
{Menten}, K.~M., {Reid}, M.~J., {Forbrich}, J., \& {Brunthaler}, A. 2007, \aap,
  474, 515

\bibitem[{{Moeckel} \& {Goddi}(2012)}]{Moeckel2012}
{Moeckel}, N., \& {Goddi}, C. 2012, \mnras, 419, 1390

\bibitem[{{Niederhofer} {et~al.}(2012){Niederhofer}, {Humphreys}, \&
  {Goddi}}]{Niederhofer2012}
{Niederhofer}, F., {Humphreys}, E.~M.~L., \& {Goddi}, C. 2012, \aap, 548, A69

\bibitem[{{Pineau des Forets} {et~al.}(1993){Pineau des Forets}, {Roueff},
  {Schilke}, \& {Flower}}]{Pineau1993}
{Pineau des Forets}, G., {Roueff}, E., {Schilke}, P., \& {Flower}, D.~R. 1993,
  \mnras, 262, 915

\bibitem[{{Plambeck} \& {Wright}(2016)}]{Plambeck2016}
{Plambeck}, R.~L., \& {Wright}, M.~C.~H. 2016, \apj, 833, 219

\bibitem[{{Plambeck} {et~al.}(2009){Plambeck}, {Wright}, {Friedel}, {Widicus
  Weaver}, {Bolatto}, {Pound}, {Woody}, {Lamb}, \& {Scott}}]{Plambeck2009}
{Plambeck}, R.~L., {Wright}, M.~C.~H., {Friedel}, D.~N., {et~al.} 2009, \apjl,
  704, L25

\bibitem[{Podio {et~al.}(2017)Podio, Codella, Lefloch, Balucani, Ceccarelli,
  Bachiller, Benedettini, Cernicharo, Faginas-Lago, Fontani, Gusdorf, \&
  Rosi}]{Podio2017}
Podio, L., Codella, C., Lefloch, B., {et~al.} 2017, Monthly Notices of the
  Royal Astronomical Society: Letters, 470, L16

\bibitem[{{Reid} {et~al.}(2007){Reid}, {Menten}, {Greenhill}, \&
  {Chandler}}]{Reid2007}
{Reid}, M.~J., {Menten}, K.~M., {Greenhill}, L.~J., \& {Chandler}, C.~J. 2007,
  \apj, 664, 950

\bibitem[{{Rodr{\'{\i}}guez} {et~al.}(2005){Rodr{\'{\i}}guez}, {Poveda},
  {Lizano}, \& {Allen}}]{Rodriguez2005}
{Rodr{\'{\i}}guez}, L.~F., {Poveda}, A., {Lizano}, S., \& {Allen}, C. 2005,
  \apjl, 627, L65

\bibitem[{{Sault} {et~al.}(1995){Sault}, {Teuben}, \& {Wright}}]{Sault1995}
{Sault}, R.~J., {Teuben}, P.~J., \& {Wright}, M.~C.~H. 1995, in ASP Conf. Ser.
  77: Astronomical Data Analysis Software and Systems IV, Vol.~4, 433.
\newblock
  \url{http://adsabs.harvard.edu/cgi-bin/nph-bib_query?bibcode=1995adass...4..433S&db_key=AST}

\bibitem[{{Schilke} {et~al.}(1997){Schilke}, {Walmsley}, {Pineau des Forets},
  \& {Flower}}]{Schilke1997}
{Schilke}, P., {Walmsley}, C.~M., {Pineau des Forets}, G., \& {Flower}, D.~R.
  1997, \aap, 321, 293

\bibitem[{Tachibana {et~al.}(2019)Tachibana, Kamizuka, Hirota, mi~Sakai, Oya,
  Takigawa, \& Yamamoto}]{Tachibana2019}
Tachibana, S., Kamizuka, T., Hirota, T., {et~al.} 2019, The Astrophysical
  Journal, 875, L29.
\newblock \url{https://doi.org/10.3847%2F2041-8213%2Fab1653}

\bibitem[{{Vaidya} \& {Goddi}(2013)}]{Vaidya2013}
{Vaidya}, B., \& {Goddi}, C. 2013, \mnras, 429, L50

\bibitem[{{Wright} \& {Plambeck}(2017)}]{wright2017}
{Wright}, M.~C.~H., \& {Plambeck}, R.~L. 2017, The Astrophysical Journal, 843,
  83

\bibitem[{{Wright} {et~al.}(1996){Wright}, {Plambeck}, \&
  {Wilner}}]{Wright1996}
{Wright}, M.~C.~H., {Plambeck}, R.~L., \& {Wilner}, D.~J. 1996, \apj, 469, 216

\bibitem[{{Zanchet} {et~al.}(2018){Zanchet}, {Roncero}, {Ag{\'u}ndez}, \&
  {Cernicharo}}]{Zanchet2018}
{Zanchet}, A., {Roncero}, O., {Ag{\'u}ndez}, M., \& {Cernicharo}, J. 2018,
  \apj, 862, 38

\end{thebibliography}


\end{document}